\documentclass[conference]{IEEEtran-date2022}
\usepackage{cite}
\usepackage{amsmath,amssymb,amsfonts}
\usepackage{algorithmic}
\usepackage{graphicx}
\usepackage{textcomp}
\usepackage{xcolor}
\usepackage{geometry}
\def\BibTeX{{\rm B\kern-.05em{\sc i\kern-.025em b}\kern-.08em
    T\kern-.1667em\lower.7ex\hbox{E}\kern-.125emX}}

\usepackage[utf8]{inputenc}
\usepackage{booktabs}
\usepackage{hyperref}

\usepackage{textcmds}
\newcommand{\mmsq}{mm\tsup{2}}
\newcommand{\BF}[1]{\textbf{#1}}

\begin{document}

\title
{
  \vspace{-0.35in}
  \LARGE Enabling Reusable Physical Design Flows \\ with Modular Flow Generators
  \vspace{-0.1in}
}

\author{
Alex Carsello, James Thomas, Ankita Nayak, Po-Han Chen,\\
Mark Horowitz, Priyanka Raina, and Christopher Torng\\[0.6em]
Stanford University, Stanford, CA
\vspace{-0.15in}
}

\maketitle



\begin{abstract}%
  Achieving high code reuse in physical design flows is challenging but increasingly necessary to build complex systems. Unfortunately, existing approaches based on parameterized Tcl generators support very limited reuse and struggle to preserve reusable code as designers customize flows for specific designs and technologies. We present a vision and framework based on modular flow generators that encapsulates coarse-grain and fine-grain reusable code in modular nodes and assembles them into complete flows. The key feature is a flow consistency and instrumentation layer embedded in Python, which supports mechanisms for rapid and early feedback on inconsistent composition. The approach gradually types the Tcl language and allows both automatic and user-annotated static assertion checks. We evaluate the design flows of successive generations of silicon prototypes designed in TSMC16, TSMC28, TSMC40, SKY130, and IBM180 technologies, showing how our approach can enable significant code reuse in future flows.
\end{abstract}


\section{Introduction}
\label{sec-intro}

Rising non-recurring engineering costs in advanced technology nodes are motivating the hardware community to adopt agile development principles and new methodologies to reduce design effort. Code reuse is particularly important to reduce the effort to build complex physical design flows.
The physical design community has been slow to adopt agile principles for a few key reasons.
First, physical design is culturally characterized by the ``one big release'' operating model with high stakes and strict annual schedules. Opportunities for code reuse disappear quickly as risk-averse teams customize scripts aggressively for their specific design, technology, and vendor libraries.
Existing approaches offered by commercial EDA vendors typically exploit reuse by leveraging parameterized Tcl templates and generators to create initial design- and technology-agnostic flows~\cite{synopsys-solvnet-web, wang-hammer-isqed2020}. These flows enable efficient reuse until a need arises for which no parameter exists. As flows are inevitably customized, these frameworks do not support propagating reusable code to different designs and technologies.
Second, the Tcl language continues to dominate commercial EDA toolflows, but it lacks language features that can help compose reusable code from different sources (e.g., introspection, gradual typing). Furthermore, modern machine learning CAD solutions are emerging that may not leverage Tcl at
all but must still compose with existing flows~\cite{lin-dreamplace-tcad2020, zhang-grannite-dac2020}. Future physical design flows seeking to reduce design effort must aggressively preserve reusable code as codebases are specialized while supporting a heterogeneous mixture of Tcl and non-Tcl code.
%


\begin{figure}[t]

  \centering
  \includegraphics[width=0.95\columnwidth]{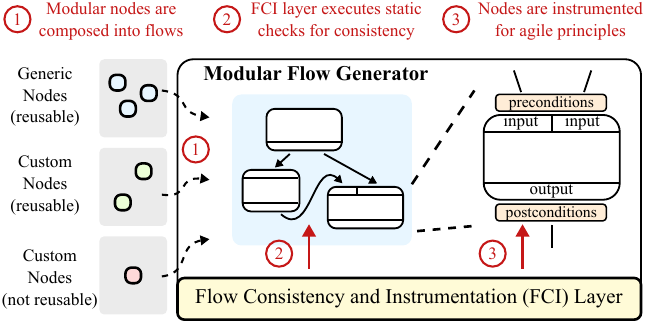}

  \caption{\textbf{A Modular Approach to Physical Design Flows} --
    Traditional Tcl-based scripts make code difficult to reuse. Modularity
    allows for reuse, and flow assembly in a high-level language (e.g.,
    Python) enables new language features and the chance to augment
    flows with agile mechanisms.
  }

  \label{fig-intro-overview}
  \vspace{-0.15in}

\end{figure}

This paper explores a vision and framework to enable reusable physical design flows based on \textit{modular flow generators} coupled with a \textit{flow consistency and instrumentation (FCI) layer} embedded in Python. Unlike existing parameterized Tcl templates and generators, the goal of a modular flow generator is not to emit Tcl but to provide the necessary abstractions to compose and reuse code. Figure~\ref{fig-intro-overview} shows how a modular flow generator composes modular nodes from both generic sources and custom sources (i.e., design- or technology-specific) in a graph that represents the assembled flow.
Since nodes from different projects can be inconsistent with each other, we introduce a Python-embedded FCI layer that provides mechanisms for both automatic and user-annotated static assertion checks across a distributed code base.
The layer can also instrument each node with dynamic assertion checks (for data-dependent use cases) and  add/remove edges in the graph to allow teams to share pre-built nodes.
While it might seem impossible to reuse code which has become design- or technology-specific, we show it can be done by refactoring the code to separate design intent (expressed with formal property checks) from the implementation. When custom nodes are then reused in a different context, the FCI layer statically executes code fragments to verify that these properties still hold.

Our work contributes
(1) \textit{mflowgen}, an open-source (m)odular (flow) (gen)erator with a flow consistency and instrumentation layer and mechanisms to reuse code across different designs and technologies, as well as rapid and early feedback on inconsistent composition;
(2) a common reusable library of over forty technology- and design-agnostic modular nodes for both commercial and open-source tool flows\footnote{
https://github.com/mflowgen/mflowgen
};
and (3) a detailed evaluation of physical design flows for silicon prototypes in TSMC16, TSMC28, TSMC40, SKY130, and IBM180 technologies, demonstrating the potential for significant code reuse in future flows.
%


\section{System Goals}
\label{sec-goals}

In this section, we overview the overarching design goals and principles that motivate our approach for physical design flows to maximize the potential for significantly reducing design effort.

\BF{Goal~1: Must achieve \textit{significant} code reuse} --
Complex physical design flows require a tremendous effort to build. Building a \textit{different but similarly complex design} will again require a similar effort. As a result, meaningfully reducing design effort will likely require that most of the physical design flow be reused. Three key design principles follow from this requirement to achieve significant code reuse. First, it is important to capture not only coarse-grain code reuse like most existing approaches (e.g., synthesis, place, route)~\cite{synopsys-solvnet-web, wang-hammer-isqed2020}, but also fine-grain reuse (e.g., glue scripts, reporting and analysis, generator wrappers). Second, we must support a mechanism to \textit{tweak} reusable code since small changes should not preclude reuse. Third, the friction to design for reuse must be low to encourage the widespread adoption necessary for success.

\BF{Goal~2: Composition must support code from different designs and different technologies} --
Physical design flows are aggressively specialized for specific designs and technologies (recall the ``one big release'' culture), and there is no avoiding this fundamental need. However, design-specific flow code can feasibly be reused across technologies (e.g., a tile-based array floorplan). Technology-specific flow code (e.g., DFM) can similarly be reused in neighboring blocks of the same design. Two key design principles follow from the requirement to support such reuse. First, our approach must support a mechanism for checking composability and consistency and a shared language for expressing requirements. Second, we must \textit{require a static code analysis approach} because code fragments in a physical design flow are distributed across tools and files and \textit{not in memory at the same time}.

\BF{Goal~3: Feedback on inconsistent composition must be both rapid and early} --
Physical design flows have very long runtimes, with RTL-to-GDS iterations often consuming days of compute on powerful server farms. As a result, dynamic runtime checks that only trigger an error when the control flow reaches problematic code will quickly result in a periodically buggy flow that breaks trust with physical designers, as well as trust in a reusable approach. Similarly, checks that are late in the flow require waiting a long time before they fire.
We make the key observation that it is reasonable and possible to separate two aspects of flows: (1) running the tools to physically construct the design, (2) running the tools to evaluate variables which turn out to be inconsistent in a composed flow. We hypothesize that generating feedback on inconsistent composition does not require the former, and we need not pay the runtime penalty.
The key takeaway that follows reinforces a static code analysis approach and formal property checks, which enables rapid and early feedback without running the actual tools.

The remainder of this work describes a concrete realization of these key design principles necessary to achieve our system goals.


\section{Modular Flow Generators}
\label{sec-system}

Our system goals motivate a flexible modular node abstraction capable of capturing both coarse- and fine-grain opportunities for reuse. Specific examples of captured code reuse may include a bump routing methodology for flip chip packages, an approach for design for manufacture (DFM) structures, adding power domains, ECO timing fixes, or a hierarchical power distribution strategy.

\subsection{Modular Node Abstraction}

The schema for the modular node abstraction is shown in Figure~\ref{fig-system-modular-nodes} and represents a function signature with file-based inputs and outputs. A node is self-contained such that the commands are executable once inputs are provided, and can contain internal scripts that may access any parameters defined in the node configuration.
Modular nodes must differ from traditional software functions because physical design depends heavily on files (e.g., netlists, databases, cell libraries).
The example synthesis node takes an input technology package and RTL design and outputs the synthesized gate-level netlist. The graph visualization shows how edges propagate files to and from the node.
There is no built-in support to ensure that nodes produce the expected results. Section~\ref{sec-python} will explore static and dynamic assertion checks for stronger guarantees.

\subsection{Categorization of Nodes that Capture Reuse}


\begin{figure}[t]

  \centering
  \includegraphics[width=0.95\columnwidth]{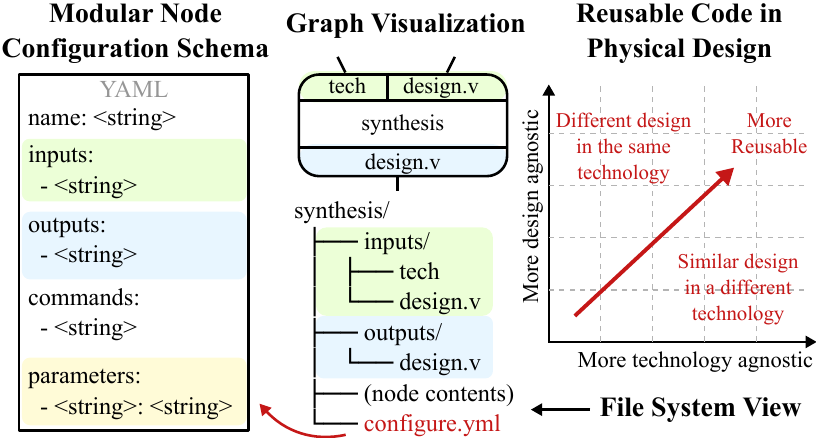}

  \caption{\textbf{Modular Node Abstraction} --
    Schema for a modular node configuration file. The ``name'' field is required. Inputs/outputs are file-based. Commands execute as shell code. A node with outputs, but no inputs or commands, acts as a vendor package. A node with inputs and commands, but no outputs, acts as an analysis node. A node with inputs, outputs, as well as commands can execute transformations.
    %
    %
  %
  }

  \label{fig-system-modular-nodes}
  \vspace{-0.1in}

\end{figure}

Figure~\ref{fig-system-modular-nodes} also shows how reusability in physical design is affected across two axes for technology-agnostic or design-agnostic code. Modular nodes that are agnostic to both are most reusable and oftentimes least performant, but many reusable code blocks have little impact on performance (e.g., converting libs-to-db).
Code agnostic along only one axis can be challenging but feasible to reuse (e.g., design-specific array-based floorplan, technology-specific DFM tasks). Code in the lower-left region has no opportunity for reuse.

Representing the upper-right region, we have built a common library of technology- and design-agnostic modular nodes as described in Table~\ref{tbl-common-lib}, which supports a wide range of common functions and can be assembled into basic flows that are functional out-of-the-box in modern technologies (see Section~\ref{sec-eval}). This capability is similar to existing work~\cite{synopsys-solvnet-web, wang-hammer-isqed2020} but is designed with our system vision.
Each common library node is parameterized (e.g., hierarchy flattening, clock gating, target slack), technology-independent (e.g., distances are multiples of metal track pitch), may be swappable between vendors (e.g., synthesis with Cadence Genus, Synopsys DC, or open-source yosys), and can be replaced entirely or decomposed into finer-grain nodes to more precisely capture reuse.


\begin{table}[t]

  \caption{Common Library of Modular Nodes}

  \centering
  \setlength{\tabcolsep}{4pt}
  \begin{tabular}{l|r|l}
    \BF{Base Tool}                   & \BF{\#} & \BF{Description of Modular Nodes}                \\
    \toprule
    Cadence Genus                    &  2      & Synthesis, generate post-pnr lib                 \\
    \midrule
    Cadence Innovus                  &  14     & Init, place, cts, route, postroute, signoff,     \\
                                     &         & post-pnr ecos, foundation flow setup,            \\
                                     &         & hold-fixing nodes, power grid setup              \\
    \midrule
    Cadence Pegasus                  &  4      & DRC, LVS, GDS merging, metal fill                \\
    \midrule
    Synopsys DC                      &  1      & Synthesis                                        \\
    Synopsys Formality               &  1      & Logical equivalence check                        \\
    Synopsys PT(PX)                  &  6      & Timing/power signoff, ECOs, gen lib/db,          \\
                                     &         & RTL- and gate-level power estimation             \\
    Synopsys VCS                     &  2      & RTL- and gate-level simulation, vcd2saif         \\
    \midrule
    Mentor Calibre                   &  7      & DRC, LVS, GDS merging, metal fill,               \\
                                     &         & convert verilog2spice, drawing chip art          \\
    \midrule
    Open-Source                      &  8      & Synthesis (yosys~\cite{yosys-web}), place (graywolf~\cite{graywolf-web}),             \\
                                     &         & place (RePlAcE~\cite{cheng-replace-tcad2018}), route (qrouter~\cite{qrouter-web})  \\
                                     &         & LVS (netgen~\cite{netgen-web}), DRC (magic~\cite{magic-web})                  \\
                                     &         & gds2spice and def2spice (magic) \\
    \midrule
    \BF{Total \# of Nodes} & \BF{45} & \\
    \midrule
    Open-Source                      &  2      & SkyWater 130nm~\cite{skywater130-web}, FreePDK45~\cite{freepdk45-web} \\
    Technologies                     &         & with NanGate Open Cell Library~\cite{nangate-cell-web} \\
  \end{tabular}
  \label{tbl-common-lib}
  \vspace{-0.15in}

\end{table}

\begin{figure}[t]

  \centering
  \includegraphics[width=\columnwidth]{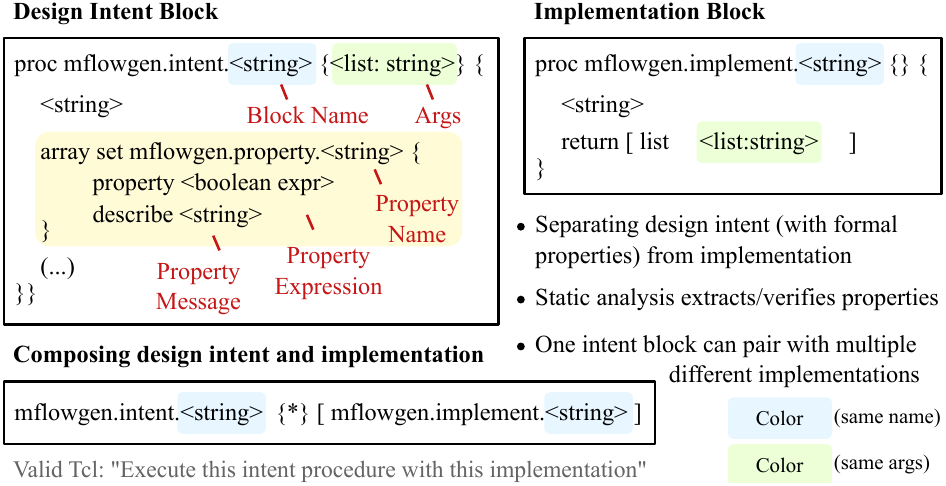}

  \caption{\textbf{Schema for Design Intent-Implementation Split}}

  \label{fig-system-intent-impl-split}
  \vspace{-0.1in}

\end{figure}

\begin{figure}[t]

  \centering
  \includegraphics[width=0.95\columnwidth]{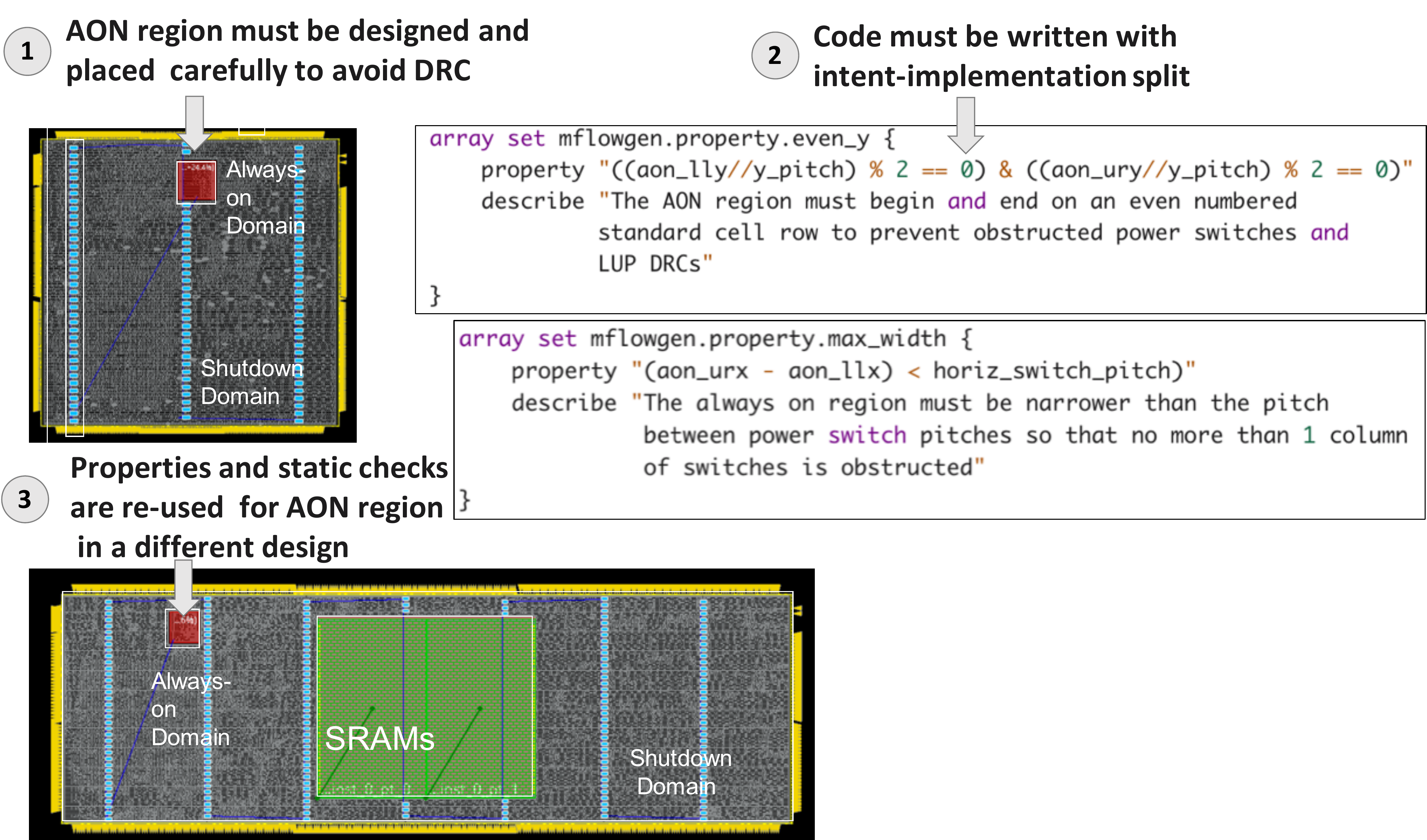}

  \caption{\textbf{Static Assertion Check Example} --
    By rewriting code with an intent-implementation split and a set of properties, the placement methodology for the always-on region within the power domain is reused across multiple designs.
  }

  \label{fig-python-static-check-aon}
  \vspace{-0.1in}

\end{figure}

Representing the upper-left and lower-right regions along each axis, Figures~\ref{fig-system-intent-impl-split} and~\ref{fig-python-static-check-aon} show how we expose reusable design intent in a way amenable to static code analysis in a Tcl context. The example shows power domains implemented for a design in 16nm. Arbitrarily choosing the placement and dimensions of the always-on power domain region can easily (and surprisingly) violate DRC when interacting with power switch columns. Traditional code can obfuscate intent and allow mistakes, resulting in DRC violations (hours later) that must be root-caused. Instead, we demonstrate a \textbf{design intent-implementation split} which allows formal properties annotated by a designer in the intent block to be evaluated with static assertion checks. The example shows how two specific properties are reused in different designs in the same technology.

This schema captures opportunities to reuse flow code by formalizing designer intent. The mechanics to analyze these code fragments across a distributed codebase are described in Section~\ref{sec-python}.

\subsection{Flow Assembly}

We allow \textit{programmatically} connecting modular nodes into graphs that represent assembled flows using a Python-embedded domain-specific language (DSL).
The DSL supports a basic graph data structure (e.g., APIs for \texttt{add\_node} and \texttt{connect}) and can add or modify parameters in each node.
This approach satisfies Goal~1 from Section~\ref{sec-goals} by providing an environment to rapidly assemble coarse-grain and fine-grain code fragments using the modular node abstraction (in contrast to existing approaches built from locked coarse-grain steps that are more difficult to modify).

A modular flow generator enables physical designer engineers to productively assemble flows of varying complexity including basic flows for initial prototyping and partial flows for test. In academia, simple teaching flows can be assembled from common library nodes and individual nodes can be incrementally swapped or added for educational purposes.
The Python DSL also opens opportunities for graph transformations, for example unrolling a loop and sweeping a parameter (e.g., clock period) for design-space exploration.


\section{Flow Consistency \& Instrumentation}
\label{sec-python}

Modular flow generators on their own provide few guarantees about the functionality of nodes and their composition, especially when nodes originate from different sources.
An otherwise reusable node for power domains may specify a specific power switch standard cell, for example, which is not composable in a different technology. The goal of the FCI layer is to enable rapid detection of inconsistencies and to provide stronger guarantees on node functionality.
Many inconsistencies can be found at run time with errors in the tools. However, tool spin times are long, and there is no guarantee how long after composition the bugs will surface. Our approach instead pulls these checks forward by running static program analysis to detect inconsistencies at graph elaboration time.

\begin{figure}[t]

  \centering
  \includegraphics[width=0.95\columnwidth]{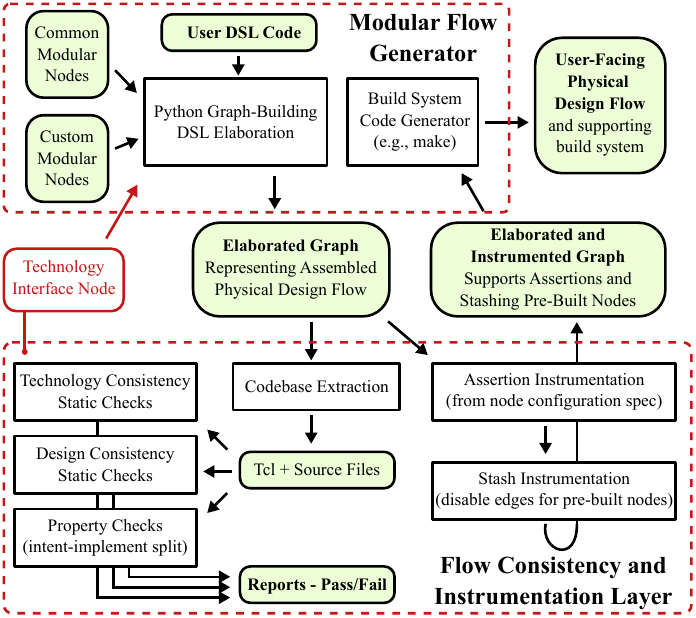}

  \caption{\textbf{Complete Toolflow Block Diagram} --
    A modular flow generator assembles the physical design flow from a user specification that composes modular nodes. The graph is checked for flow consistency and instrumented to enable agile principles before producing the final flow as a build system.
  }

  \label{fig-toolflow}
  \vspace{-0.1in}

\end{figure}

Figure~\ref{fig-toolflow} shows the complete toolflow that composes modular flow generators with a Python-embedded FCI layer. The user-specified graph is first elaborated by the flow generator, resulting in a detailed in-memory graph model representing the assembled physical design flow and the source locations for each modular node in the file system. The FCI layer introspects the graph model and gathers all Tcl and source files for static analysis. Static checking then flags potential inconsistencies.


\subsection{Consistency Checks}

The property checks described in Figure~\ref{fig-system-intent-impl-split} are lifted as boolean expressions and evaluated after running the implementation split as an emitted Tcl fragment. Aside from these checks, we also provide mechanisms for both automatic and user-annotated static assertion checks in an extensible way.

Our framework can be used to \textit{gradually type} the Tcl language, which has only a string type (not very useful).
For example, we can extend the framework for technology consistency checks (see FCI layer's modular technology interface node in Figure~\ref{fig-toolflow}). An annotation for \texttt{mflowgen.enum.stdcell(INV\_X1)} can indicate an \textit{enum} construct that is automatically defined from a routine that reads the LEF macros in the technology and flags invalid standard cells. Similar checks can be built to check other parameters (e.g., pitches, delay units, valid metal layers).
Extending the framework for design consistency checks is similar but does not require technology access. For example when constructing a tile array, annotating each tile's floorplan height with \texttt{mflowgen.equality.tile\_height(\$var)} checks that all equality blocks with name ``tile\_height'' have expressions ``var'' resolving to the same value (the FCI layer emits and executes Tcl fragments).
All annotations are embedded in Tcl as pass-through procs and do not change the Tcl semantics.


\begin{figure}[t]

  \centering
  \includegraphics[width=\columnwidth]{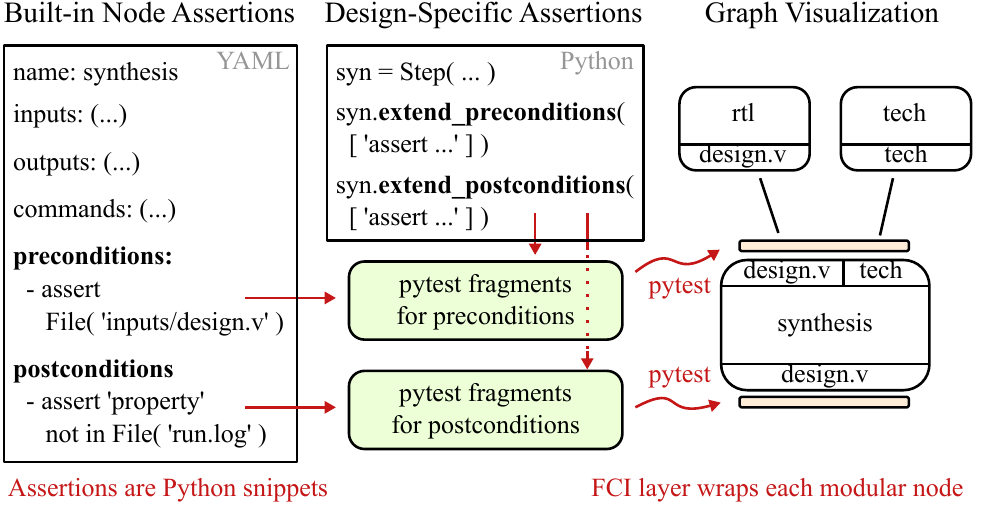}

  \caption{\textbf{FCI Layer Instrumentation for Assertions} --
    The modular abstraction allows for built-in assertions in each node as
    well as design-specific assertions.
  }

  \label{fig-agile-assertions}

\end{figure}


\begin{figure}[t]

  \centering
  \includegraphics[width=\columnwidth]{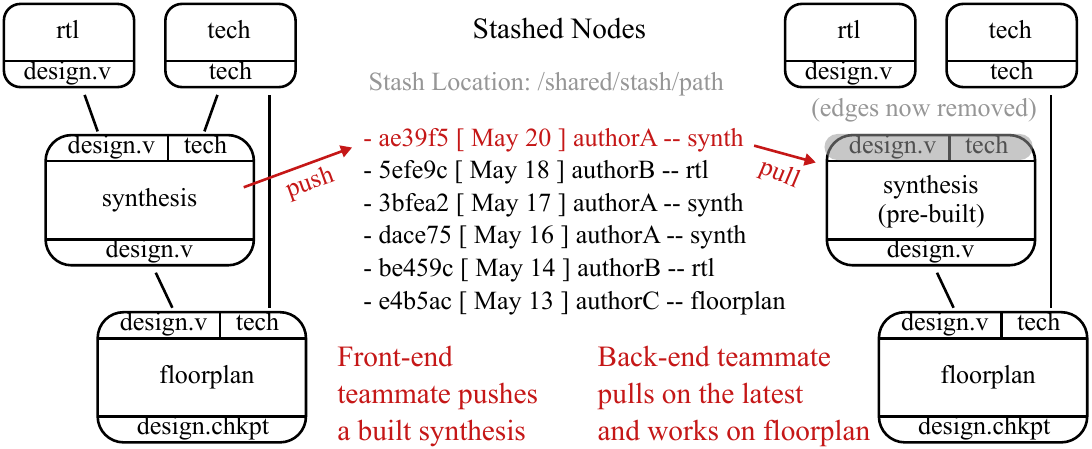}

  \caption{\textbf{FCI Layer Instrumentation for Pre-Built Nodes} --
    Pre-built modular nodes can be stashed as vendor packages and shared across a team. The FCI layer transforms the graph to remove input dependency edges.
  }

  \label{fig-agile-stash}
  \vspace{-0.1in}

\end{figure}

\subsection{Flow Instrumentation}


\begin{table*}[t]

  \caption{Chips Built and Fabricated with Modular Flow Generators}

  \centering
  \setlength{\tabcolsep}{2pt}
  \begin{tabular}{l|c|c|c|c|c|c}
                                       & \BF{DenseAccel16}  & \BF{MiniCGRA}  & \BF{CryptoAccel}        & \BF{DenseAccelRRAM}      & \BF{RVMulticore} & \BF{BaseSynch} \\ 
    \midrule            
    Application Domain                 & Image              & Image          & Cryptography            & Machine                  & General          & Wireless  \\
                                       & Processing / ML    & Processing     &                         & Learning                 & Purpose          &           \\
    \midrule                      
    Technology                         & TSMC16             & SKY130         & SKY130                  & TSMC40                   & TSMC28           & IBM180    \\
    \hline                               
    Area                               & 25\mmsq            & 10\mmsq        & 10\mmsq                 & 29.2\mmsq                & 1.25\mmsq        & 2.31\mmsq \\
    \hline                             
    Max Frequency                      & 750 MHz            & 60 MHz         & 325 MHz                 & 200 MHz                  & 500 MHz          & 20 MHz    \\
    \hline                               
    Voltage                            & 0.9 V              & 1.8 V          & 1.8 V                   & 1.1 V                    & 0.9 V            & 1.8 V     \\
    \hline                              
    Power                              & 0.5-1 W            & -              & -                       & 126 mW                   & ~10s of mW       & -         \\
    \hline                              
    Number of Cores                    & 384 PE, 128 MEM    & 24 PE, 8 MEM   & 1                       & 256 PE                   & 4                & 2         \\
    \hline                               
    On-Chip Memory                     & 4.6 MB             & 8 KB           & None                    & 2MB/0.5MB RRAM/SRAM      & 64 KB            & None      \\
    \hline                    
    Memory Levels                      & 3                  & 2              & None                    & 3                        & 1                & None      \\
    \hline                               
    Has Off-Chip Memory                & Yes                & No             & No                      & Yes                      & No               & No        \\
    \hline                    
    Physical Hierarchies               & 3                  & 2              & 1                       & 2                        & 1                & 1         \\
    \hline                               
    Multiple Power Domains             & Yes                & No             & No                      & Yes                      & No               & No        \\
    \hline                               
    Number of Clock Domains            & 3                  & 1              & 1                       & 4                        & 1                & 4         \\
    \hline                               
    \% of Codebase (LoC) reused         & 30\%               & 58\%           & 94\%                    & over 80\%                & 86\%             & 84\%      \\
    from common library                 &                    &                &                         &                          &                  &           \\
    \hline                               
    \% of Codebase (LoC) reused         & 50\%               & 24\%           & First design            & First design             & First design     & First design \\
    from previous designs               &                    &                &                         &                          &                  &           \\
    \hline                               
    Months to tapeout                   & 6                  & 2.5            & 2.5                     & 6                        & 2                & 1.5       \\
    \hline                             
    Static check runtime                & 2.2 sec            & 0.8 sec        & 0.2 sec                 & 0.6 sec                  & $<$1 sec           & $<$1 sec    \\
    \hline                             
  \end{tabular}
  \label{tbl-chips-v2}
  \vspace{-0.05in}

\end{table*}

Figure~\ref{fig-toolflow} illustrates how the FCI layer can instrument all modular nodes with additional functionality.
The modular node configuration schema can specify optional \textit{dynamic assertion checks}, which the FCI layer inserts before and after each node. These cover scenarios where a desired check is data-dependent (e.g., parsing and flagging unexpectedly poor-quality results, or specific problematic errors in logs). Pre- and post-conditions are Python snippets.
Figure~\ref{fig-agile-assertions} shows how these checks can be built into a node or extended in the Python graph model, where pre- and post-conditions simply appear as Python lists.
Assertions are run with pytest~\cite{pytest-web}, a full-featured software testing tool.
In addition, modular nodes create natural checkpoints that can be shared across a team. Figure~\ref{fig-agile-stash} shows how each node can be \textit{stashed} into a shared team space from which other team members can pull pre-built nodes into their graphs. On a stash pull, the FCI layer transforms the graph in-place to break input dependency edges, resulting in a static vendor package that simply supply outputs and is never re-built, \textit{regardless of the built state of prior nodes} (unlike Makefiles).


\section{Evaluation}
\label{sec-eval}

We apply our approach to build silicon prototypes in multiple technologies to evaluate code reuse. Our primary emphasis and indication of success will be (1) achieving significant code reuse for custom code to build 2nd+ generation designs, because we expect existing frameworks~\cite{synopsys-solvnet-web, wang-hammer-isqed2020} to perform similarly in supporting 1st generation designs, and (2) the speed of static assertion checks running on large codebases for complex physical design flows.


\begin{figure}[t]

  \centering
  \includegraphics[width=0.8\columnwidth]{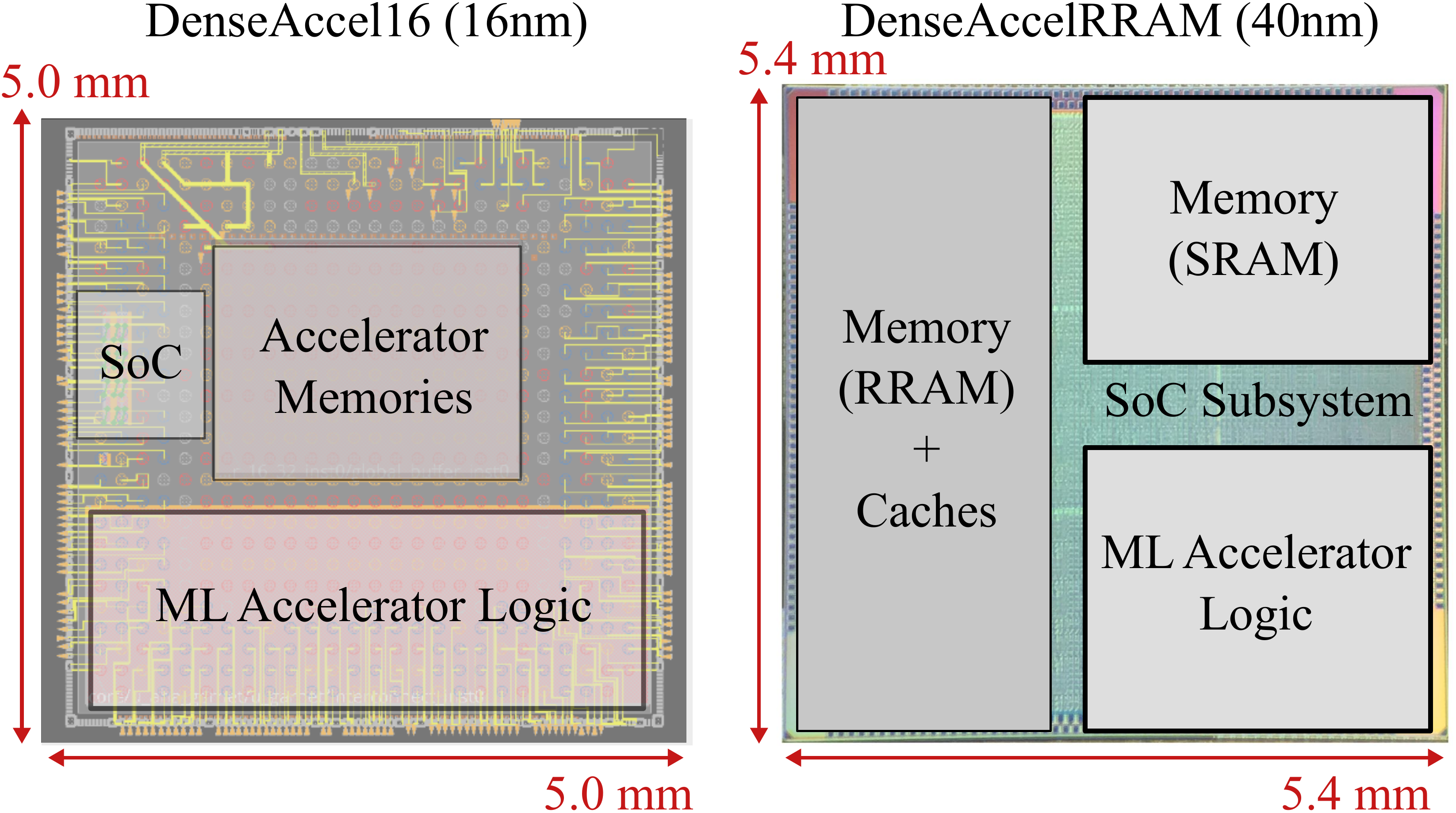}

  \caption{\textbf{Two accelerator-based SoC silicon prototypes}}

  \label{fig-case-studies-chips}
  \vspace{-0.1in}

\end{figure}

Table~\ref{tbl-chips-v2} lists the high-level specifications of each chip and the features that made physical design challenging.
At a glance, total code reuse for each of the six chips was very high (totals of 80\%+ lines of code reused), with 2nd+ generation designs achieving good coverage with code from previous designs. All chips were completed with very short timelines of less than six months. Figure~\ref{fig-case-studies-chips} shows layouts for DenseAccel16 and DenseAccelRRAM before tapeout.


\subsection{Evaluating Static Assertion Checks}

\begin{figure}[t]

  \centering
  \includegraphics[width=0.98\columnwidth]{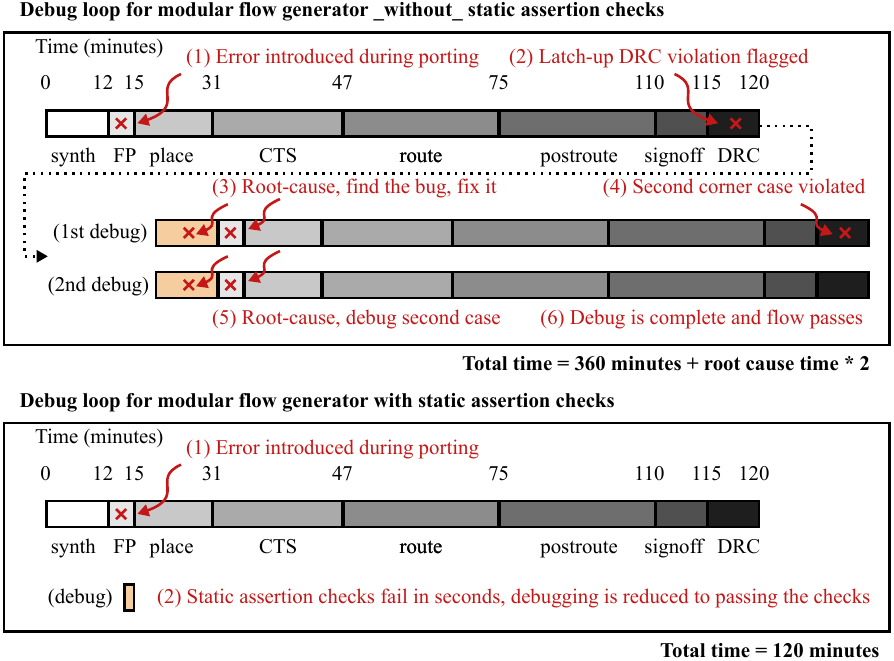}

  \caption{\textbf{Debug loop with and without static assertion checks for power domains case study in Figure~\ref{fig-python-static-check-aon}}.
  }

  \label{fig-results-debug-loop}
  \vspace{-0.1in}

\end{figure}

We evaluate the benefit of static assertion checks with a breakdown of flow tool spin times in DenseAccel16 in Figure~\ref{fig-results-debug-loop}. The entire synthesis-to-DRC flow completes in about 120 minutes on our servers for the particular sub-block under design. The case study corresponds to Figure~\ref{fig-python-static-check-aon}, where the goal is to take an implementation of power domains from the processing element tile design (square layout) and port to the memory tile design (rectangular layout).

\textbf{Impact on debug loop} -- In the baseline flow without static assertion checks, the engineer must run the entire flow through DRC to discover a latch-up DRC violation (two hours later in this design, but potentially far longer in larger designs). The engineer then root-causes the bug (orange bar split in timeline) which we annotate as a ``root cause time'' variable but can span minutes, hours, or days. After spending effort to debug and fix this bug in isolation, which includes understanding the DRC report, the purpose of all code statements, and filtering lines for blame, the engineer attempts the full flow again only to discover a second bug (in our case, the second property in Figure~\ref{fig-python-static-check-aon}). This lengthy debug loop can repeat multiple times.
In the second flow that includes static assertion checks, the engineer no longer needs to rediscover the \textit{design intent} that in the former case was lost over time in the original code being ported. Because static assertion checks formalize these requirements, and because the modular flow generator executes these checks statically in a few seconds at graph elaboration time (before any physical design tools are run), the figure timeline shows that the entire debug loop becomes far less painful.


\textbf{Static assertion check runtimes} -- Table~\ref{tbl-chips-v2} quantifies the FCI layer runtime to inspect the entire codebase for each chip. We sum the times for hierarchical sub-designs, with each number collected over five trials on a 2.4GHz Quad-Core Intel Core i5-8279U laptop-class CPU. The static check runtimes are quick, ranging from 0.2--2.2s for the largest codebase. We ran a study scaling up to 1000 intent-implement blocks, which still completes in under 20s.

\subsection{Benefit of Modularity}

We also discuss a case study for evaluating the modular flow generator approach itself over existing approaches based on parameterized Tcl generators~\cite{synopsys-solvnet-web, wang-hammer-isqed2020}, which encourage implementing custom features in injected Tcl hooks (e.g., pre- and post- each step). This results in monolithic Tcl scripts that tangle many concerns (e.g., power domains, floorplan, chip IO, DFM) as opposed to our modular node approach which has one node per concern.

The \textbf{design} is DenseAccel16 and two previous iterations of the same design (less complex) in the same 16nm technology and metallization. All three iterations (we will call them DenseAccel16-1 through -3) included processing element tiles with power domains, but the first was monolithic Tcl and the second/third were built as modular flow generator nodes. The \textbf{time to port power domains code from DenseAccel16-1 to DenseAccel16-2 was two months, while the time from DenseAccel16-2 to DenseAccel16-3 was two days, both for a single student}.
We attribute the time difference to the tangling of features in monolithic Tcl scripts, requiring our designers to spend months understanding every line of code and gathering all lines related to power domains into one place (and then debugging all of this in a loop). In contrast, moving from the second to the third design was far simpler because a single node captured all code related only to power domains, and this node was designed as a vendor package supplying code fragments across the flow.

\subsection{Reuse in MiniCGRA}


\begin{figure}[t]

  \centering
  \includegraphics[width=0.99\columnwidth]{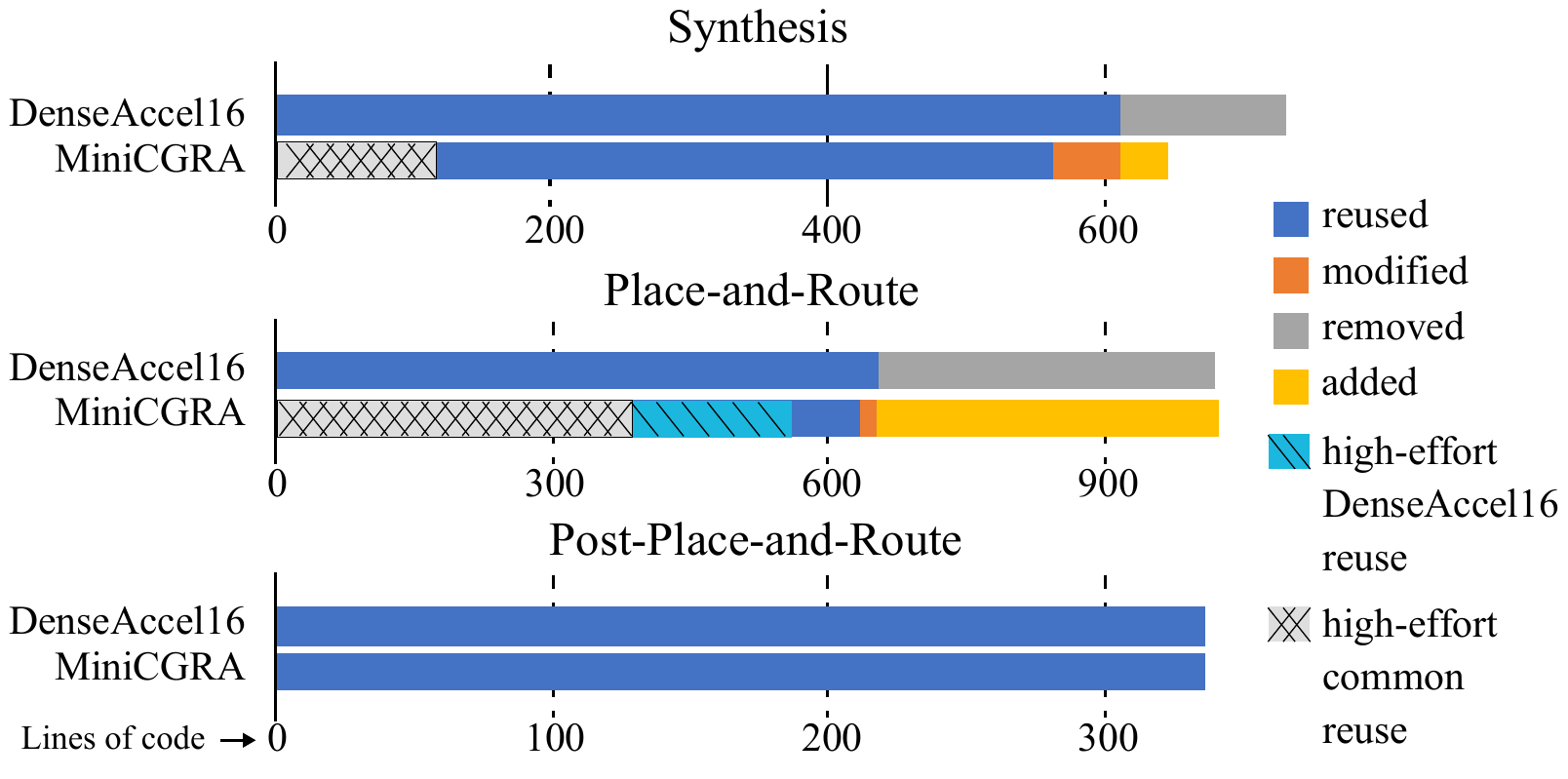}

  \caption{\textbf{Mini CGRA Reuse} --
    Lines of code reused to build a small CGRA in SKY130 technology, sourcing code from both common library nodes and the codebase of the previous CGRA in TSMC16 technology.
  }

  \label{fig-results-reuse-mini-cgra}

\end{figure}

MiniCGRA is a coarse-grain reconfigurable array (CGRA) derived from a portion of DenseAccel16, but implemented in an open-source SKY130 technology~\cite{skywater130-web}.
Figure~\ref{fig-results-reuse-mini-cgra} and Table~\ref{tbl-chips-v2} illustrate the detailed breakdown of code reused, modified, added, and removed from the common library as well as from custom design-specific (but technology-agnostic) code from the DenseAccel16 codebase.
Our breakdown also splits bars to visualize high-effort code designed for reuse, which are carefully written to derive important values from the technology library instead of hard-coding specific numbers. For example, the tile array placement automatically detects the width and height of both compute and memory tiles and lays them out in a grid to abut pins. This coding effort significantly increases reuse to a total of 82\% for the final codebase. In summary, a modular approach allowed custom work to be cleanly inserted or replaced in the form of new modular nodes, mitigating the challenges of working on a large physical design codebase.
%


\section{Conclusion}
\label{sec-conclusion}

While physical designers would like modular and reusable flows, today's tools, design approaches, and deadlines lead to flows being tuned aggressively and destroying reusable code.
This paper presents a system vision and framework that can help physical design flows maximize the potential for significantly reducing design effort. Our modular flow generator approach provides the abstractions for composing coarse-grain and fine-grain code fragments and provides mechanisms for embedding checks and extending Tcl (i.e., a framework for gradual typing) to ensure that these properties hold.
We developed a concrete implementation of our modular flow generator approach and fabricated silicon prototypes in multiple technologies to demonstrate the potential for significantly reducing design effort in future flows.




\bibliographystyle{ieeetr}
\bibliography{main}

\begin{thebibliography}{10}

\bibitem{synopsys-solvnet-web}
``Synopsys reference methodology retrieval system.'' Online Webpage, 2021
  (accessed Sep 18, 2021).
\newblock \url{http://solvnet.synopsys.com/rmgen}.

\bibitem{wang-hammer-isqed2020}
E.~Wang, C.~Schmidt, A.~Izraelevitz, J.~Wright, B.~Nikoli\'{c}, E.~Alon, and
  J.~Bachrach, ``A methodology for reusable physical design.,'' {\em Int'l
  Symp. on Quality Electronic Design (ISQED)}, Mar 2020.

\bibitem{lin-dreamplace-tcad2020}
Y.~Lin, Z.~Jiang, J.~Gu, W.~Li, S.~Dhar, H.~Ren, B.~Khailany, and D.~Z. Pan,
  ``Dreamplace: Deep learning toolkit-enabled gpu acceleration for modern vlsi
  placement,'' {\em IEEE Trans. on Computer-Aided Design of Integrated Circuits
  and Systems (TCAD)}, Jun 2020.

\bibitem{zhang-grannite-dac2020}
Y.~Zhang, H.~Ren, and B.~Khailany, ``Grannite: Graph neural network inference
  for transferable power estimation,'' {\em Design Automation Conf. (DAC)}, Jul
  2020.

\bibitem{yosys-web}
``Yosys open synthesis suite.'' Online Webpage, 2020 (accessed Nov 23, 2020).
\newblock \url{http://www.clifford.at/yosys}.

\bibitem{graywolf-web}
``Graywolf.'' Online Webpage, 2020 (accessed Nov 23, 2020).
\newblock \url{https://github.com/rubund/graywolf}.

\bibitem{cheng-replace-tcad2018}
C.-K. Cheng, A.~B. Kahng, I.~Kang, and L.~Wang, ``Replace: Advancing solution
  quality and routability validation in global placement,'' {\em IEEE Trans. on
  Computer-Aided Design of Integrated Circuits and Systems (TCAD)}, Jul 2018.

\bibitem{qrouter-web}
``Qrouter.'' Online Webpage, 2020 (accessed Nov 23, 2020).
\newblock \url{http://opencircuitdesign.com/qrouter}.

\bibitem{netgen-web}
``Netgen.'' Online Webpage, 2020 (accessed Nov 23, 2020).
\newblock \url{http://opencircuitdesign.com/netgen}.

\bibitem{magic-web}
``Magic.'' Online Webpage, 2020 (accessed Nov 23, 2020).
\newblock \url{https://github.com/RTimothyEdwards/magic}.

\bibitem{skywater130-web}
``Skywater open source pdk.'' Online Webpage, 2021 (accessed Sep 18, 2021).
\newblock \url{https://github.com/google/skywater-pdk}.

\bibitem{freepdk45-web}
``Freepdk45.'' Online Webpage, 2020 (accessed Nov 23, 2020).
\newblock \url{https://www.eda.ncsu.edu/wiki/FreePDK45:Contents}.

\bibitem{nangate-cell-web}
``Silvaco nangate open cell library.'' Online Webpage, 2020 (accessed Nov 23,
  2020).
\newblock \url{https://si2.org/open-cell-library}.

\bibitem{pytest-web}
``Pytest.'' Online Webpage, 2014 (accessed Oct 1, 2014).
\newblock \url{http://www.pytest.org}.

\end{thebibliography}

\end{document}